\documentclass{icrc29}
\usepackage{graphicx,amssymb,amsmath,times}
\begin{document}
\title{Cosmic ray electron acceleration in the remnant of nova GK Persei?}
\author[Kantharia et al.] {N.G. Kantharia$^a$, G.C.Anupama$^b$, Prasad Subramanian$^c$ \\
         $^a$ National Centre for Radio Astrophysics (TIFR), 
            Post Bag 3, Ganeshkhind, Pune 411 007, India \\
         $^b$ Indian Institute of Astrophysics, II Block Koramangala, Bangalore - 560 034, 
            India \\  
         $^c$ Inter-University Centre for Astronomy and Astrophysics, Post Bag 4,
               Ganeshkhind, Pune 411 007, India \\ }
\presenter{Presenter: N.G.Kantharia (ngk@ncra.tifr.res.in) \
ID code: ind-kantharia-NG-abs1-og14-poster} 

\maketitle

\begin{abstract}
Radio observations of the nova remnant of GK Per around 1985 (epoch 1)
have shown it to have a spectral index of $ \sim -0.4$ ($S \propto \nu^{\alpha}$) at frequencies
$<1.4$ GHz and a spectral index of $-0.7$ at frequencies
$>1.4$ GHz \cite{Seaquist}, \cite{Biermann}.
This spectrum is either due to secondary electrons from p-p interactions \cite{Biermann} or
due to the presence of at least two electron
populations which dominate the emission at different frequencies.
Our radio observations \cite{Anupama} of
the remnant of GK Persei using the Giant Metrewave Radio Telescope (GMRT) and
archival Very Large Array (VLA) data show that the nova remnant 
has a low frequency ($<1.3$ GHz) spectral index $\alpha \sim -0.85$ (in 2003; epoch 3) and 
a high frequency spectral index of  $\alpha \sim -0.7$ (1997; epoch 2). 
Moreover the higher frequencies show an annual secular decrease of
$\sim 2.1 \%$ between epochs 1 and 2 whereas there is no change at 0.33 GHz.
This suggests a significant evolution of the spectrum over 20 years. In this paper, we
discuss a possible physical basis for this spectral evolution based on a comparison of the
synchrotron loss and electron acceleration timescales.
\end{abstract}

\section{Introduction}
GK Persei is a classical nova system that had a nova outburst in 1901. 
Classical novae are a type of cataclysmic variable objects which are
characterized by a significant mass loss ($\sim 10^{-5}-10^{-4} M_\odot$)
during an outburst which are infrequent (once in about $10^4-10^5$ years).  
GK Per consists of a white dwarf primary accreting matter from
an evolved K2 subgiant secondary \cite{Anupama1} and the system
is embedded in an extended bipolar nebula \cite{Bode}, \cite{Seaquist},
probably a leftover from its earlier avataar and
which continues to influence the evolution of the nova remnant from
the 1901 outburst.   

The nova remnant has been extensively studied in the optical \cite{Anupama}
(and references therein).  Radio observations at 1.5 GHz and 4.9 GHz by \cite{Reynolds}
discovered the presence of a non-thermal, polarized radio remnant, with a
morphology resembling the optical remnant. A detailed multifrequency study of the nova remnant
and the environs of GK Per were made by \cite{Seaquist} which included radio
continuum study at frequencies between 0.151 and 4.86 GHz, in which they
found that the radio spectrum had a low frequency turnover around
1 GHz, with the spectrum being flat at the lower frequencies. Biermann et al. \cite{Biermann} 
have reanalysed some of the Seaquist et al. \cite{Seaquist} data and find that the low frequency
spectrum has $\sim -0.4$.  They suggest that the origin of this spectrum
lies in the p-p interactions.  The radio
emission and the optical images of the remnant indicate that the nova ejecta
are interacting primarily with the ambient medium located to the southwest of
the nova. The optical nebular remnant shows a deceleration in this direction.

Anupama \& Kantharia (2005) recently studied the nova remnant in the optical
using the 2-m Himalayan Chandra Telescope (HCT) and in the 
radio continuum at 0.33, 0.61 and 1.28 GHz using the GMRT (2003; epoch 3). 
They also used VLA archival data at 1.49 and 4.86 GHz obtained in 1997 (epoch 2).
Anupama \& Kantharia \cite{Anupama} find an annual secular decrease
of about 2.1\% in the flux density at 1.49 and 4.86 GHz. They also find that there is
a break in the spectrum near 1 GHz with the spectrum getting steeper at the lower
frequencies, in conflict with the Seaquist et al \cite{Seaquist} (epoch 1) result.  
Here, we try to explain the results of Anupama \& Kantharia (2005) and
comment on the contribution of GK Per-like novae to the total cosmic ray content.  

\section{The radio emission}
\begin{figure}[h]
\begin{center}
\includegraphics*[width=4.5cm,angle=-90]{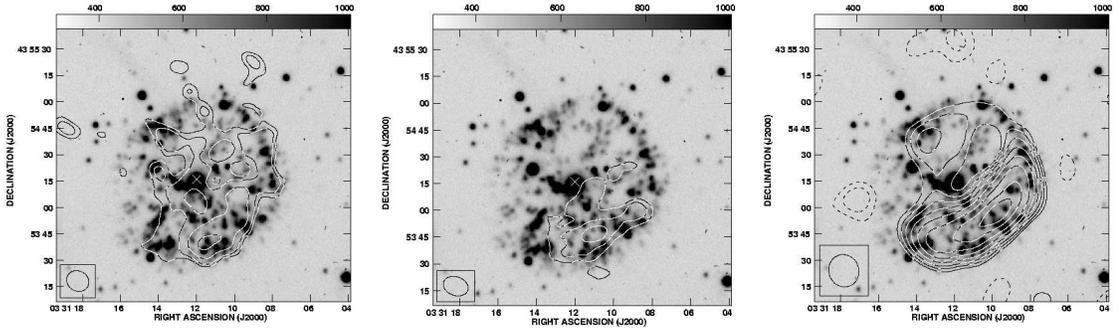}
\caption{\label {fig1} GMRT radio images (contours) of GK Per nova remnant shell at
0.33 GHz (left) and 0.61 GHz (middle) and VLA 1.4 GHz (right) superposed
on the [NII] optical image (grey scale) from HCT (From \cite{Anupama}).}
\end{center}
\end{figure}

Fig. \ref{fig1} shows the GMRT images superposed on the optical [NII] HCT image. 
The optical and radio shells are of similar extent. 
The low-brightness radio emission extends to the north-east
where Anupama \& Kantharia (2005) report enhanced density.
The peak emission at 0.33 and 0.61 GHz appears to arise from a location 
that is different from that of the peak emission at 1.4 GHz.  
This might be indicating the presence of at least
two dominant electron populations in the nova remnant or different
physical processes dominating at the different frequencies.  The spectrum
of the high frequency peak is slightly flatter compared to the 
low frequency peak probably indicating local acceleration. 

The radio spectrum of the nebular remnant of GK Per 
which includes data from Seaquist et al. \cite{Seaquist} and
Biermann et al. \cite{Biermann} who revised the flux estimates given by Seaquist et al. at
0.33 and 0.61 GHz is presented in Fig. \ref{fig3}(a).  For convenience we refer
to the data from Seaquist et al. and Biermann et al. taken around 1985 as epoch 1,
the archival VLA data from 1997 as epoch 2 and our data in 2003 as epoch 3 in this paper.
From the figure, it is clearly seen that
the flux at all frequencies has decreased with time, except at 0.33 GHz,
where the flux has remained unchanged.  The decrease in flux density at 1.49 and 4.8 GHz
can be explained by a secular decrease of 2.1\% per year \cite{Anupama}
which has left the spectral index unchanged at $-0.7$  
whereas at 0.6 GHz, the decrease is about 1.7\% per year.  
The object which had a spectral index of $-0.7$ between
1.4 and 4.8 GHz and a flatter index of $\sim -0.4$ at frequencies lower than 1.4 GHz
(Seaquist et al. 1989, Biermann et al 1996) seems to have evolved in the last 18 years
giving a dramatic steepening of the low frequency 
spectrum to an index of -0.85 for frequencies lower than 1.3 GHz \cite{Anupama}
These observations raise questions such as what could be the underlying physical
reason for the two spectra which we discuss in the next section.

\section{Spectral evolution and electron acceleration}
In epoch 1, the spectrum had a spectral index $-0.7$ at frequencies above 1.4 GHz
and an index of $-0.4$ at lower frequencies.  The most convincing
explanation we could find in the literature was due to Biermann et al. \cite{Biermann}.
Assuming that the electrons responsible for the radio emission were secondary
products of p-p collisions, they derived a high frequency spectral index of $-0.7$ 
and a low frequency index of $+1/3$.  It may be noted that the low frequency
spectral index that Biermann et al. \cite{Biermann} predict does not match the observed value. 

As mentioned earlier, we find that the spectrum above 1.4 GHz has undergone
a secular decrease between epoch 1 and epoch 2.  There are no measurements below
1.4 GHz during epoch 2 and no measurements above 1.4 GHz during epoch 3. 
We find that the low frequency spectrum (below 1.3 GHz)  has a considerably
steeper spectral index ($-0.85$) during epoch 3.  Assuming that the high
frequency spectrum continues to follow the same secular decrease observed
between epochs 1 and 2, we have the following picture for the composite
spectrum during epoch 3: a high frequency ($>1.4$ GHz) spectrum with a spectral index of
$-0.7$ and a low frequency ($<1.3$ GHz) spectrum with an index of $-0.85$. 
Comparing the composite spectrum at epoch 1 with this putative composite
spectrum at epoch 3, we find that the low frequency spectrum has steepened
while the high frequency spectrum has retained its spectral index while undergoing
a secular decrease.

\begin{figure}[h]
\begin{center}
\includegraphics*[width=7cm]{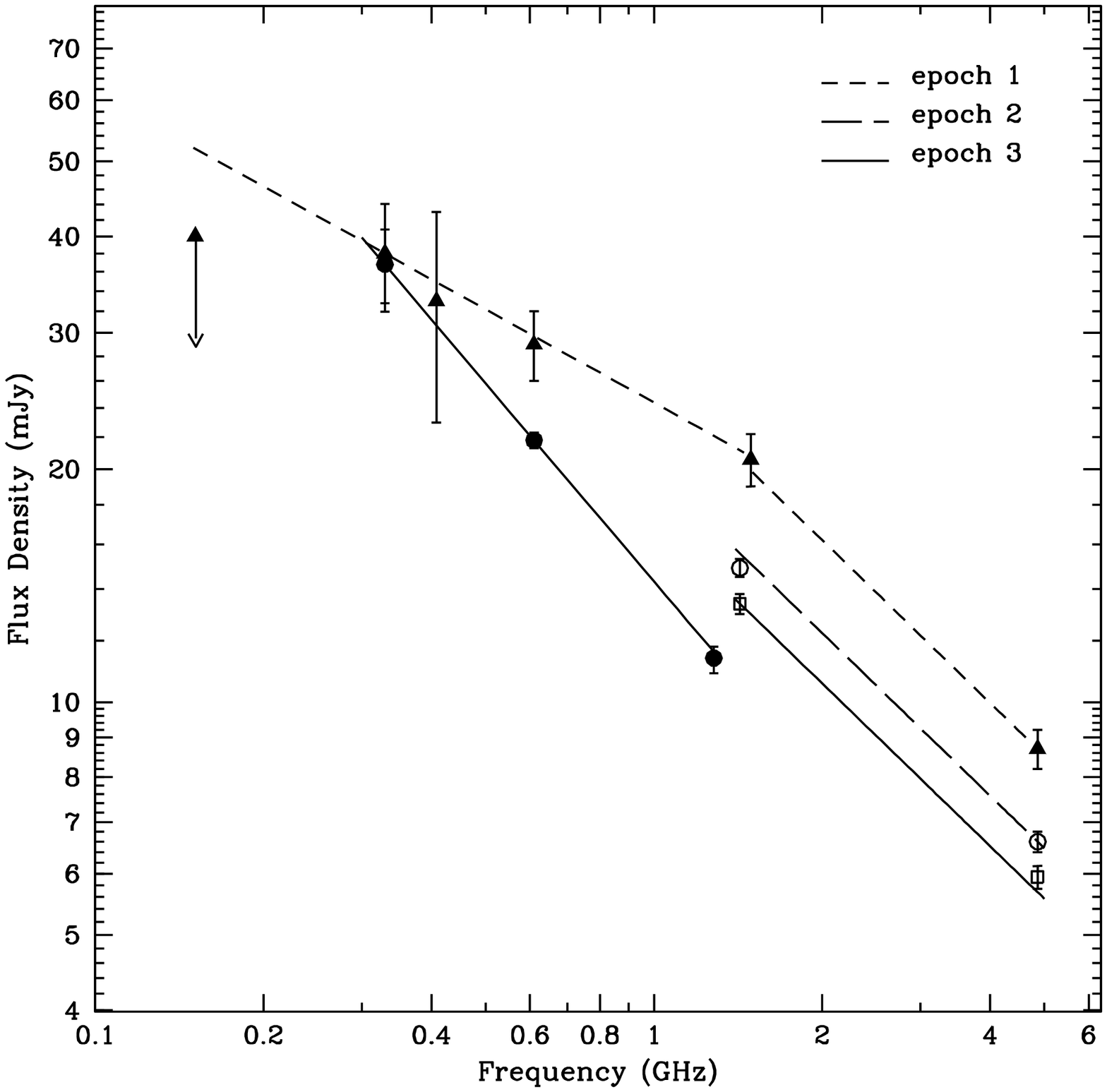}
\includegraphics*[width=7cm]{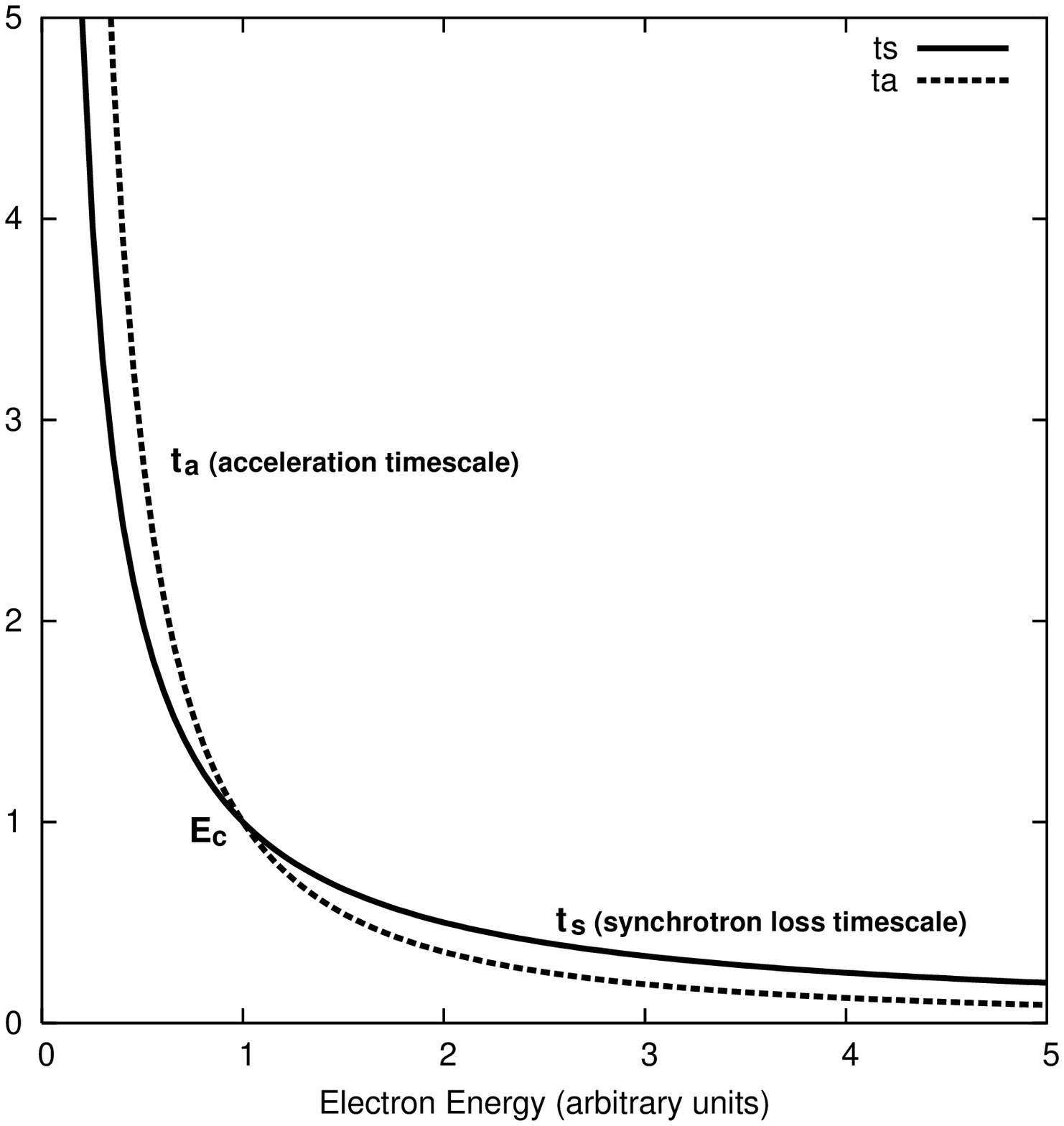}
\caption{\label {fig3} (a) Radio spectrum of emission from the nova remnant of GK Per
at different epochs (data from \cite{Anupama}, \cite{Biermann} \cite{Seaquist}).
The lines connecting the points at frequencies above 1.4 GHz 
are for a spectral index of $-0.7$ whereas the line connecting epoch 3 data is for $\alpha=-0.85$
and the line connecting the low frequency points of epoch 1 is for $\alpha=-0.4$. (b) 
A schematic showing the timescales of acceleration and synchrotron losses
which can explain the observed spectrum in epoch 3.  Thus, the electrons at low energy
should have larger losses whereas at the high energies, the acceleration
process should balance (or be slightly larger than) the losses.} 
\end{center}
\end{figure}

We now attempt to analyse the effects of possible acceleration, 
synchrotron losses and adiabatic losses suffered by the electrons that could
explain such an evolution of the composite spectrum. At high frequencies, 
if the acceleration timescale $t_a$ is close to (or slightly smaller than) the synchrotron loss timescale $t_s$ 
the spectrum will not undergo changes in its spectral index and will merely
shift downwards (i.e., undergo a secular decrease) as a result of adiabatic expansion. 
At low frequencies, however, if $t_s < t_a$, it would lead to a steepening of the spectrum over time.
Since the synchrotron loss timescale $t_s \propto 1/E$ where E is the electron energy,
the argument outlined above suggests a form for the acceleration timescale 
\begin{equation}
t_{a} \propto E^{-\beta}\, , \,\,\,\,\,\,\, \beta \lesssim 1
\label{e1}
\end{equation}
as shown in Fig \ref{fig3}(b).  We denote the energy at which $t_a$ and $t_s$ 
intersect by $E_c$.  The break frequency between the high and low energy
spectra corresponds to the critical frequency of an electron with
energy $E_c$.  Fig \ref{fig3}(b) suggests that acceleration balances (or slightly dominates) 
over losses at electron energies $> E_c$ while losses dominate for electron
energies $< E_c$.  This could be a possible explanation for the evolution
of the composite spectrum from epoch 1 to epoch 3. 

The time evolution of the electron distribution function $f$ due to 
acceleration is commonly described by 
the following diffusion equation (e.g., Subramanian, Becker \& Kazanas 1999)
\begin{equation}
\frac{\partial f}{\partial t} = -\frac{1}{E^{2}}\,\frac{\partial}{\partial E}\,\biggl (-E^{2}\,{\cal D}\,\frac{\partial f}{\partial E} \biggr ) \, .
\label{e2}
\end{equation} 
This equation describes the evolution of the distribution function as a result of diffusion in electron energy E.
The acceleration timescale arising from such a mechanism can be written as 
(appendix A of \cite{Subramanian}) 
\begin{equation}
t_{a}^{-1} = \frac{1}{E^{3}}\,\frac{\partial}{\partial E}\,\biggl (E^{2}\,{\cal D} \biggr ) \, .
\label{e3}
\end{equation}

If the energy dependence of the diffusion co-efficient is of the form
${\cal D} \propto E^{\xi}$,
the requirement on the acceleration timescale expressed in Eq~(\ref{e1}), when combined with Eq~(\ref{e3})
implies that $3 \leq \xi $.
The physical mechanism accelerating the electrons must yield a diffusion co-efficient ${\cal D}$ which satisfies this requirement.

\section{Contribution to galactic cosmic rays}
The mininum energy in relativistic particles required to explain the observed synchrotron
emission from GK Per using equipartition arguments comes out to be
$10^{34} \eta^{4/7}$ Joules
where $\eta$ is the ratio of energy contained in protons and electrons.
This is fairly insignificant compared to the energy contained in particles in
supernova remnants like Cas~A ($\sim 2\times10^{41} \eta^{4/7}$ Joules) which
are significant contributors to the cosmic ray flux.
Thus, although the nova rate of 30 yr$^{-1}$ \cite{Shafter} is fairly high
compared to the supernova rate of 0.02 yr$^{-1}$ \cite{Allen},
the energy released per classical nova is insignificant and hence contributes
little to the total cosmic ray flux.

\noindent
{\bf Acknowledgements}\\
We thank the staff of the
GMRT that made these observations possible. GMRT is run by the
National Centre for Radio Astrophysics of the Tata Institute of Fundamental
Research. This work has made use of The NRAO Data Archives. The National Radio
Astronomy Observatory is a facility of the National Science Foundation (U.S.A.)
operated under cooperative agreement by Associated Universities, Inc.
We thank the staff of IAO, Hanle and CREST, Hosakote, for their support during
observations. The facilities at IAO and CREST are operated by the Indian
Institute of Astrophysics, Bangalore.


\begin{thebibliography}{99}

\bibitem{Allen}
C. W. Allen, in Astrophysical Quantities, Pub: The Athlone Press, University of
London (1976) 

\bibitem{Anupama}
G. C. Anupama \& N. G. Kantharia, Astron. \& Astrophy., 435, 167 (2005)

\bibitem{Anupama1}
G. C. Anupama \&  T. P. Prabhu, Mon Not of Roy Astron Soc, 263, 335 (1993).

\bibitem{Bode}
M. F. Bode, E. R. Seaquist \& A. Evans, Mon Not of Roy Astron Soc, 228, 217 (1987).

\bibitem{Biermann}
P. L. Biermann, R. G. Strom, H. Falcke, Astron. \& Astrophy.,302, 429 (1995).

\bibitem{Reynolds}
S. P. Reynolds, \& R. A. Chevalier, Astrophy J., 281, L33 (1984).

\bibitem{Seaquist}
E. R. Seaquist et al., Astrophy J., 344, 805 (1989).

\bibitem{Shafter}
A. Shafter in Classical Nova Explosions. ed. M. Hernanz \& J. Jose, AIP Conf. Ser. 637, 462 

\bibitem{Subramanian}
P. Subramanian, P. A. Becker, D. Kazanas, Astrophy J., 523, 203 (1999).

\end{thebibliography}
\end{document}